\documentclass[prl,twocolumn,floatfix,showpacs,amsmath,amssymb]{revtex4}
\usepackage{graphicx}

\newcommand\beq{\begin{equation}}
\newcommand\eeq{\end{equation}}
\newcommand\bea{\begin{eqnarray}}
\newcommand\eea{\end{eqnarray}}
\newcommand\non{\nonumber}
\newcommand\bib{\bibitem}

\begin{document}



\title{\bf Effect of strong correlation on the study of renormalization
group flow diagram for Kondo effect in a interacting quantum wire. 
}
\author{Sujit Sarkar}
\affiliation{\it PoornaPrajna Institute of Scientific Research, 
4 Sadashivanagar, Bangalore 5600 80, India.
}

\date{\today}

\begin{abstract}
We present the study of Kondo effect in an interacting quantum wire.
We mainly emphasis the effect of strong electronic correlations in 
the study of renormalization group flow diagram
and the stability analysis of fixed points for both magnetic and
nonmagnetic impurities. We observe that the behavior of the system
is either in the single channel or 
in the two channel Kondo effect depending
on the initial values of coupling constants and strong correlations. 
\end{abstract}

\pacs{: 73.63.Nm, 72.15.Qm, 71.10.Pm }
\maketitle


\section{I. Introduction}
One dimensional interacting electron systems are interesting in their own
right. In one dimension, interaction between the electrons has large
impact on the low-energy properties of the electron liquids. The basic
physics of electron liquid is changing from the Fermi liquid (FL) to
the Luttinger liquid (LL) physics. Such an one dimensional LL can be realized 
in a very narrow quantum wire.
In this work, we are interested in the study of Kondo 
effect in a quantum wire, where both magnetic and nonmagnetic impurities
are present. Kondo effect is one of the interesting phenomena of
correlated electronic systems since its discovery \cite{kon, noz1, noz2}. 
This effect arises
from the exchange interactions between an impurity spin and an electron
gas in three dimensions. In this case interaction changes from weak to strong
coupling as one decrease the temperature or energy scale. At T=0 the impurity
spin is completely screened by the conduction electrons and the basic physics 
can be understood from the local FL \cite{kon, noz1, noz2}. It was also observedthat when 
the channel ($N$) exceeds twice of the impurity spin the system 
is governed by the
physics of non-Fermi liquid theory \cite{andr, aff}.\\
In the conventional theory of Kondo effect, people have worked on an 
independent 
electron picture because in three dimensions the interacting electron system
can be described by the noninteracting quasiparticle (FL). However this scenario
is drastically changing in one dimension when the basic physics is govern by the LL.
The Kondo effect in Luttinger liquid was first discussed by Lee and Toner \cite{lee}.
Furusaki 
and Nagaosa \cite{nag} have derived the scaling equations for the Kondo couplings in the
weak coupling regime by using the poor-man's scaling method, 
which preserves the
$SU(2)$ symmetry. They predicted that at low temperature the system is governed
by the strong coupling fixed point, where the impurity spin is screened completely.
They observed an important difference with conventional Kondo physics is that 
Kondo coupling flows to the strong coupling regime for 
both antiferromagnetic and
ferromagnetic coupling because the local backward scattering 
potential is relevant
perturbation in the LL. But we shall see in our study that strong correlation
effect has not taken into account properly, i.e., the weak coupling RG flow 
diagram is the same for noninteracting, repulsive and attractive LL.
The problem of a spin-1/2 magnetic impurity in a LL has also 
been largely studied
in the literature \cite{john,durga,hur, egg,furu,ravi,meden}.
It is well known that the repulsive
interaction due to nonmagnetic impurity breaks the wire at the impurity 
site and treats the
residual tunneling through the barrier as a perturbation.
Here we study the two (dual) models that describe the impurity.
The plan of the work is
the following: In section II, 
we present the weak coupling renormalization group flow
diagram with necessary physical analysis for a one dimensional interacting
electron system with only magnetic impurity. In sec. III, we present 
weak-coupling RG 
flow diagram of an one dimensional system with both magnetic and nonmagnetic impurities
are present. The section IV is devoted for conclusions.
\section{II. Weak coupling renormalization group study of a 
quantum wire with only magnetic impurity}
Here we consider a magnetic impurity of spin-1/2 at the origin of one dimensional interacting
systems. It is well known that away from half-filling the basic physics of this type of system is
governed by the spin sector of the Hamiltonian. We only consider the exchange coupling between the
impurity spin and conduction electrons as follows  
\bea
H_{1} & = & \frac{J_1}{2} S. [ {{\psi}_{R \alpha}}^{\dagger} {\sigma}_{\alpha \beta} 
{{\psi}_{R \alpha}}~ +~{{\psi}_{L \alpha}}^{\dagger} {\sigma}_{\alpha \beta}
{{\psi}_{L \alpha}}] \non\\
& & +~ \frac{J_2}{2} S. [ {{\psi}_{R \alpha}}^{\dagger} {\sigma}_{\alpha \beta} 
{{\psi}_{L \alpha}}~ +~{{\psi}_{L \alpha}}^{\dagger} {\sigma}_{\alpha \beta} {{\psi}_{R \alpha}}],
\label{ham1}
\eea
where $J_1$ and $J_2$ are respectively the forward and backward Kondo scattering coupling,
$\sigma$ is the pauli matrix. ${\psi}_{R \alpha}$ and ${\psi}_{L \alpha}$  
are respectively the fermionic right and left mover field operators with spin $\alpha$.
The field operators of right and left going electrons with spin $\sigma$ are,
$ {\psi}_{R \sigma} (x) = \frac{{\eta}_{\sigma}}{\sqrt{2 \pi a}} e^{i {\phi}_{R} (x) }$, 
$ {\psi}_{L \sigma} (x) = \frac{{\eta}_{\sigma}}{\sqrt{2 \pi a}} e^{- i {\phi}_{L} (x) }$,
where $\phi$ is the bosonic field. 
$\eta$ is the Klein factor that preserve the anticommutivity of fermionic field.
During our continuum field theoretical calculations,
we consider the following relations of the bosonic fields as
$ {\phi}_c (x) ~=~ \frac{1}{2} [ {\phi}_{L \uparrow} ~+~ {\phi}_{L \downarrow}~+~  
{\phi}_{R \uparrow} ~+~  {\phi}_{R \downarrow}] $,
$ {\theta}_c (x) ~=~ \frac{1}{2} [ {\phi}_{L \uparrow} ~+~ {\phi}_{L \downarrow}~-~  
{\phi}_{R \uparrow} ~-~  {\phi}_{R \downarrow}] $,
$ {\phi}_s (x) ~=~ \frac{1}{2} [ {\phi}_{L \uparrow} ~-~ {\phi}_{L \downarrow}~+~  
{\phi}_{R \uparrow} ~-~  {\phi}_{R \downarrow}] $ 
$ {\theta}_s (x) ~=~ \frac{1}{2} [ {\phi}_{L \uparrow} ~-~ {\phi}_{L \downarrow}~-~  
{\phi}_{R \uparrow} ~+~  {\phi}_{R \downarrow}] $.
Finally we get the bosonized Hamiltonian
\bea
H ~&=&~ \frac{J_1 S^-}{4 \pi a}~[ e^{i ({\theta}_s (0) -{\phi}_s (0))}
~+~ e^{-i ({\theta}_s (0) +{\phi}_s (0))}] ~+~H.C \non\\ 
& &- \frac{J_1 S_z}{\pi} {\partial}_x {\phi} (x) +
  \frac{J_2 S^-}{2 \pi a}~cos( {\phi}_c (0)~-~{\theta}_s (0) ) \non\\ 
& & ~+~  \frac{J_2 S^+}{2 \pi a}~cos( {\phi}_c (0)~+~{\theta}_s (0) )  \non\\ 
& &- \frac{J_2 S_z}{\pi a} sin ({\phi}_c (0))  sin ({\phi}_s (0)) . 
\eea
The scaling dimension for the forward scattering terms 
and 
backward scattering terms are respectively 1 and $\frac{1}{2} (1 + K_c)$. 
So the 
backward scattering term is relevant for $K_c < 1$ 
, where $K_c$ is the LL parameter of the charge sector.
One can derive the 
renormalization group equation for one loop order by using poor-man's scaling method.
The RG equations are the following \cite{nag,furu}: 
\bea
\frac{d J_1}{dl} ~&=&~ \frac{1}{2 \pi v_F}~( {J_1}^2 ~+~ {J_2}^2 ) ,
\non \\
\frac{d J_2}{dl} ~&=&~ (~ \frac{1}{2} (1- K_c) J_2 ~+~ \frac{2 J_1 J_2}{2 \pi v_F} ~)  \non \\,
\label{rg1}
\eea
These RG equations have only trivial fixed point, $({J_1}^{*} , {J_2}^{*} ) = (0,0)$.
We do the linear stability analysis to check the stability of these fixed points (FP). After
the linear stability analysis RG equations reduce to
\beq
\frac{d}{dl} A_1~= ~\frac{1}{2 \pi v_F} B_1 A_1 ,
\eeq    
where  
\[ A_1 = \left (\begin{array}{c}
       J_1\\J_2
        \end{array} \right ) \] and
\[ B_1 = \left (\begin{array}{cc}
      {J_1}^{*} & {J_2}^{*} \\
      {J_2}^{*} & \frac{(1-K_c)}{2} + {J_1}^{*}
        \end{array} \right ) \].
At the trivial fixed point, $\frac{d J_1}{dl}~=~0*J_1$ and 
$\frac{d J_2}{dl}~=~\frac{(1-K_c)}{2}*J_2$.
The equations for $J_1$ and $J_2$ will be stable and unstable when the
right hand side coefficient of $J_1$ and $J_2$ are negative and positive
respectively. The couplings are marginal when the coefficients are zero.
If we look at the next order term for the marginal case, i.e., 
$\frac{d J_1}{dl}~=~ a {J_1}^2$ ($ a >0 $), we say that FP at $J1~=~0$
is stable on the $ x<0 $ side and unstable on the $x >0$ side. We get from the
linear stability analysis of $J_2$ that the fixed point is stable when $K_c > 1$,
is marginal for $K_c~=~1$ and is unstable for $K_c <1$.\\
\begin{figure}[]
\includegraphics[height=7cm,width=9.0cm,,angle=0]{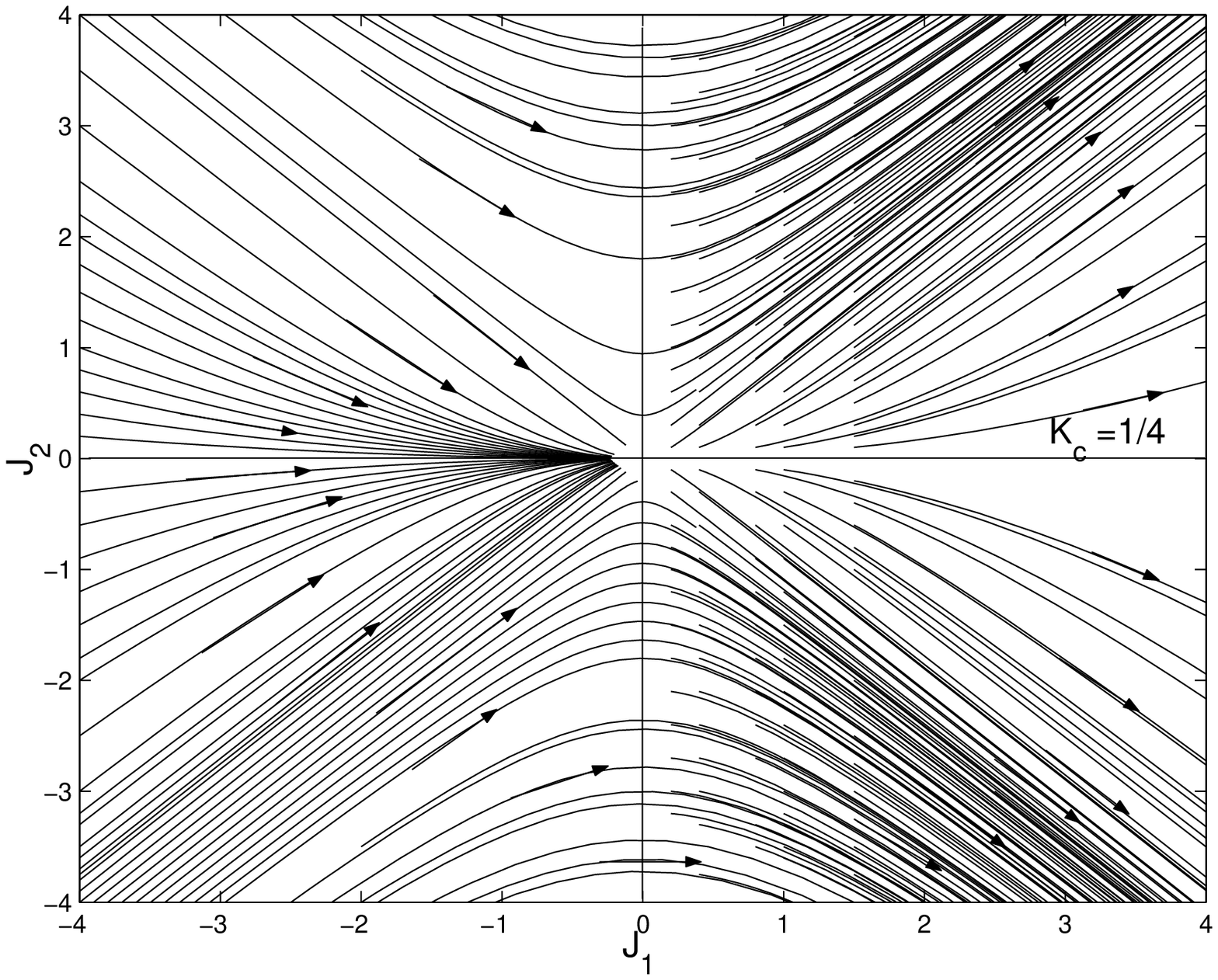}
\caption{The RG flow diagram in the $J2-J1$ plane for Eq. 3 . The solid line and arrow 
show the flow and the direction respectively, $J_1 , J_2 \ll 2 \pi v_F$.
}
\label{Fig1}
\end{figure}
In Fig. 1, we present the RG flow diagram for the repulsive quantum wire, i.e,
$K_c~=~1/4$. The Kondo couplings ($J_1$ and $J_2$) are renormalized towards strong
coupling irrespective of the couplings are antiferromagnetic or ferromagnetic
Kondo couplings. This observation is in contrast with the three dimensional 
Kondo effect. We observe from our study that the RG flow takes a large range of
initial conditions to the FP at (0,0). For all other initial conditions, we 
see that there are two directions along which the Kondo 
couplings flow to strong 
coupling $\frac{J_2}{J_1}~=~1$ and $\frac{J_2}{J_1}~=~-1/2$. This explicit
study of the phase diagram was absent in the previous studies of 
the Kondo effect in an one
dimensional interacting quantum wire \cite{nag,furu}.\\
The one-loop RG equations suggest that apart from the trivial FP
(0,0), there are other three FPs, ($ {J_2}^*, {J_1}^*$) = (+ $\infty, + \infty$),
(+ $\infty, 0$), (+ $\infty, - \infty$). Here we want to discuss the FP (+ $\infty, 0$)
. It corresponds to the two channel Kondo problem. When $J_2 =0$, the spin and 
the charge sectors are decoupled  in the bosonized Hamiltonian. The 
impurity spin is interacting with ${\phi}_s $ and ${\theta}_s $ only and
the Hamiltonian of the spin sector is equivalent to the 
two-channel Kondo problem
with the right and left going electrons correspond to two channel.\\
\begin{figure}[]
\includegraphics[height=7cm,width=9.0cm,,angle=0]{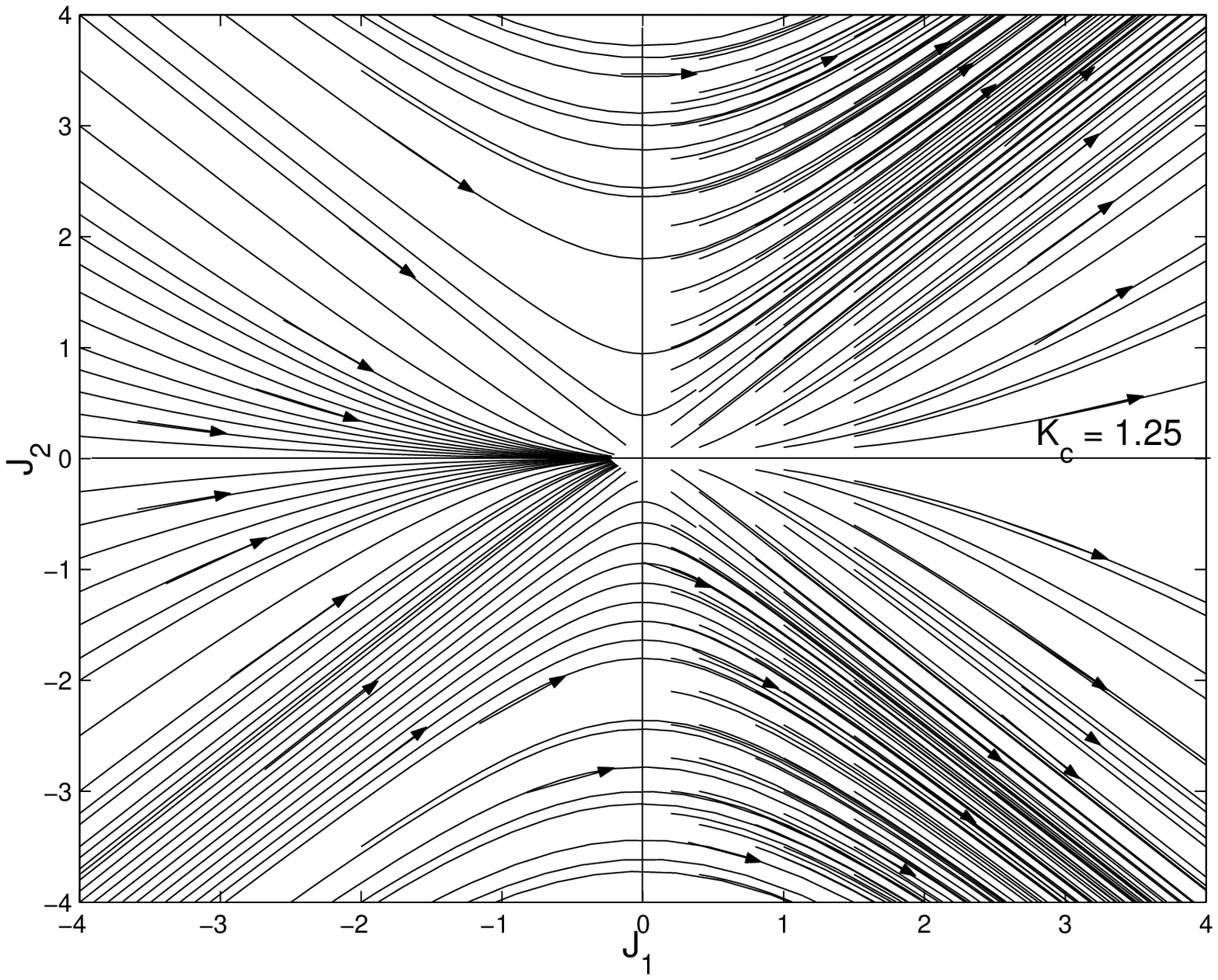}
\caption{The RG flow diagram in the $J2-J1$ plane for Eq. 3. The solid line and arrow 
show the flow and the direction respectively, $J_1 , J_2 \ll 2 \pi v_F$. }
\label{Fig2}
\end{figure}
In Fig. 2, we present the RG flow diagram for $K_c~=~3/2$, i.e, the system of 
attractive LL. We observe that the flow diagram and the two channel Kondo
regime is the same as repulsive LL ($K_c ~=~1/4$). It also reveals from our
study that noninteracting RG flow diagrams show the same behavior as the attractive and 
repulsive LL. So we conclude that the behavior of the RG flow diagram is 
the characteristic of the one dimensional system, interactions (different values 
of LL parameter) have no effect. 
This explicit study of strong electronic correlation on the study of Renormalization
Group flow diagram was absent in the previous study [7,12].
The RG equations of a quantum wire with a 
magnetic impurity
are not sufficient to show up the strong correlation effects
because there is no breaking of quantum wire. Therefore one has to
consider the presence of strong nonmagnetic scatterer, which we will discuss
in the next section.\\  

\section{Renormalization group study of a quantum wire with magnetic impurity
and nonmagnetic local potential}

Here we present the RG study of an interacting wire with magnetic 
($J_1$) and nonmagnetic ($J_2$) local potential. We study this part following 
Ref. 15. We consider the situation where the strength of local impurity potential
is larger than the antiferromagnetic exchange $J_1$. For this case one can 
first diagonalize the Hamiltonian with only local potential. 
It is well known that
for repulsive interaction due to nonmagnetic impurity
cutting the wire at the impurity site.
Here we consider the tunneling $t$ as a perturbation. The Hamiltonian of the system 
is the following,   
\bea
H_{2} & = & \frac{J_1}{2} S. \sum_{i = R,L} \sum_{\alpha , \beta} 
{{\psi}_{i \alpha}}^{\dagger} {\sigma}_{\alpha \beta} 
{{\psi}_{i \alpha}} \non\\
& & +~ \frac{J_2}{2} S. [ {{\psi}_{R \alpha}}^{\dagger} {\sigma}_{\alpha \beta} 
{{\psi}_{L \beta}}~ +~ H.C ] \non\\
& & +~ t ~ \sum_{\alpha} ( {{\psi}_{R \alpha}}^{\dagger} {{\psi}_{L \alpha}}~+~ H.C),
\label{ham2}
\eea
Where R and L correspond to the right and left side of the impurity, $S$ is the
impurity spin-1/2 operators and $\sigma$ is the spin operator of the conduction
electrons. The first term represents the exchange interaction of the leads 
with the impurity spin. The second term represents the tunneling process 
with spin-flip. The third term presents the tunneling of electrons without 
spin flip.\\
This model Hamiltonian represents the two channel Kondo model when $J_2$ and
$t$ are absent. The presence of $J_2$ and $t$ will introduce anisotropy
between two channels. In Ref. 15, the RG equation for this problem has derived.
We think that this derivation is not the complete one, i.e., there is one
more RG equation for the tunneling without spin flip. Now we present the 
derivation of that RG equation very briefly. 
One can write the boundary field operator \cite{fab} as
\beq
{\psi}_{i \uparrow / \downarrow}~=~ \frac{{\eta}_{i \uparrow / \downarrow}}
{\sqrt{ 2 \pi a}}~e^{( \frac{{\phi}_{ic}}{\sqrt{2 K_c}}~\pm~
  \frac{{\phi}_{is}}{\sqrt{2 K_s}}) } ,
\eeq
where ${\phi}_{is}~=~ \frac{1}{\sqrt{2}} ( {\phi}_{i \uparrow}~-~
 {\phi}_{i \downarrow})$ and 
${\phi}_{ic}~=~ \frac{1}{\sqrt{2}} ( {\phi}_{i \uparrow}~+~{\phi}_{i \downarrow})$. 
$K_s$ is the LL parameter of the spin sector.
Using these expressions for fermionic and bosonic fields, one can write 
the bosonized version of the tunneling without spin flip component of the
Hamiltonian as 
$$ \frac{2}{\pi a}~ cos (\frac{{\phi}_{Rc} - {\phi}_{Lc}}{\sqrt{2 K_c}}) 
cos (\frac{{\phi}_{Rs} - {\phi}_{Ls}}{\sqrt{2 K_s}}).$$ The scaling
dimension of this term is $\frac{1}{2} (\frac{1}{K_c}~+~ \frac{1}{K_s})$ and
there is no term alike $t$ from the operator product expansion of the first
two terms of the Hamiltonian. So the one-loop RG equation
is
\beq
\frac{dt}{dl}~=~ ( 1 ~-~ \frac{1}{2} ~( \frac{1}{K_c}~+~\frac{1}{K_s})~)t
\eeq
Therefor the total RG equations are
\bea
\frac{d J_1}{dl} ~&=&~ \frac{1}{2 \pi v_F}~( {J_1}^2 ~+~ {J_2}^2 ) ,
\non \\
\frac{d J_2}{dl} ~&=&~ (~ \frac{1}{2} (1- \frac{1}{K_c}) J_2
 ~+~ \frac{2 J_1 J_2}{2 \pi v_F} ~), \non \\
\frac{dt}{dl}~&=& ( 1 ~-~ \frac{1}{2} ~( \frac{1}{K_c}~+~\frac{1}{K_s})~)t
\label{rg2}
\eea
\begin{figure}[]
\includegraphics[height=7cm,width=9.0cm,,angle=0]{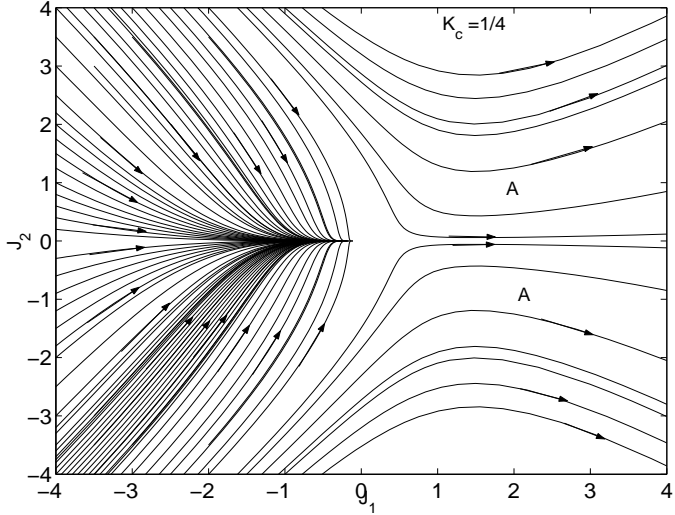}
\caption{RG flow diagram in the $J_2-J_1$ plane for Eq. 8. The solid line and arrow 
shows the flow and the direction respectively. }
\label{Fig3}
\end{figure}
Here the FPs are the trivial ones, i.e., 
$({J_1}^* , {J_2}^* , t^*)= (0,0,0)$.
We do linear stability analysis of the FPs to study the nature of the FPs.
After the linear stability analysis the RG equations are reduced to
\beq
\frac{d}{dl} A_2~= ~\frac{1}{2 \pi v_F} B_2 A_2 ,
\eeq    
where 
\[ A_2 = \left (\begin{array}{c}
       J_1\\J_2\\t
        \end{array} \right ) \] and
\[ B_2 = \left (\begin{array}{ccc}
      {J_1}^{*} & {J_2}^{*}  & 0 \\
      {J_2}^{*} & \frac{(1-K_c)}{2} + {J_1}^{*} & 0 \\
       0 & 0 & \frac{(1 -\frac{1}{K_c})}{2}
        \end{array} \right ) \].

\begin{figure}[htbp]
\includegraphics[height=7cm,width=9.0cm,,angle=0]{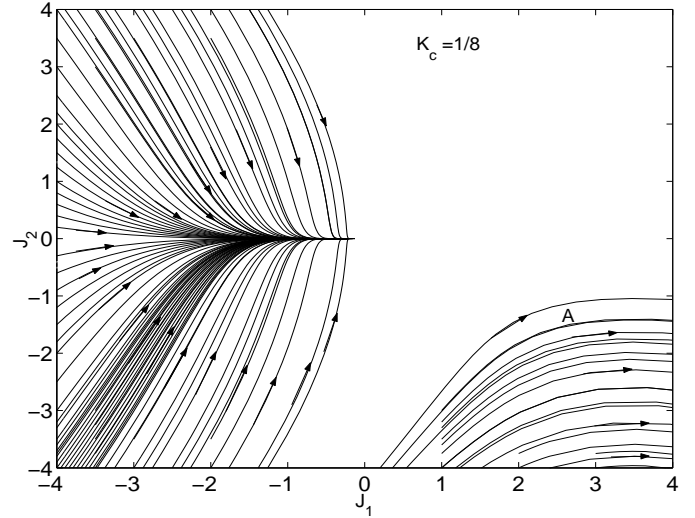}
\caption{The RG flow diagram in the $J_2  - J_1 $ plane for Eq. 8. The solid line and arrow 
show the flow and the direction respectively. }
\label{Fig4}
\end{figure}
At the trivial fixed point, $\frac{d J_1}{dl}~=~0*J_1$, 
$\frac{d J_2}{dl}~=~\frac{(1-K_c)}{2}*J_2$ and 
$\frac{d t}{dl}~=~\frac{1}{2}\frac{(1-K_c)}{2} t$.
We understand from these RG equations that the coupling $J_1$ is marginal, $J_2$ and $t$ 
will become relevant or irrelevant depending on 
whether $( 1~-~1/K_c)$ is greater than or
less than zero. As we understand from the analysis of these RG equations that 
for study of 
the $J_1$ vs. $J_2$ flow diagram, the first two RG equations are 
sufficient to draw the
phase diagram. At the large length scale, the FP is approached as 
$J_1 ~\sim~\frac{-1}{ln (T_K /T)}$ and 
$J_2 ~\sim~{(\frac{T}{T_K})}^{\frac{1 - 1/K_c}{2 \pi v_F}}$.
Here $T_K$ is the Kondo temperature. The behavior of $J_1$ and $J_2$ at
large length scale in contrast
to the behavior of FL. In this case the Eq. 8, can be solved in terms
of the linear combinations of, $J_1$ and $J_2$ such that,  
$J_1 ~-~J_2$ and $J_1 ~+~J_2$. Here the fixed point is also
at  
$({J_1}^* , {J_2}^*)= (0,0)$ with $J_1 ~\sim~\frac{-1}{ln (T_K /T)}$ and 
$J_2 ~\sim~{(\frac{T}{T_K})}^{2}$. We see that $J_2$ approaches zero faster than $J_1$
for both FL and LL but for the LL, it goes to zero much faster power of $T$. The above
analysis is valid when neither $J_1~-~J_2$ nor $J_1 ~+~ J_2$ is exactly equal to zero.
If one of them is exactly equal to zero and the other is not,
then both the couplings go as 
$\frac{1}{ln (T_K /T)}$. This explicit study was absent in Ref. [15]\\
Before we proceed further with the analysis of the RG flow diagrams, 
we first want to study 
the bosonized version of the Hamiltonian to get a better understanding of our results.
The bosonized version of this model Hamiltonian has discussed in Ref. 15. 
We have only
borrowed the final results.
$J_{1 \perp}$ (XY component of $J_1$) is always relevant and $J_{2 \perp}$ is relevant 
only if $1/2 < K_c <1$. When both $J_{1 \perp}$ and $J_{2 \perp}$ are relevant 
the low temperature FP behavior is the same as the single channel Kondo effect.
Two channel Kondo
behavior is stable for $K < 1/2$. 
We are now presenting the RG flow diagram for different values of LL
parameters.
We see from our
RG study that for $1/2 < K_c < 1$, a certain region of the parameter
space system shows the two channel and single channel 
Kondo behavior as we see in Fig. 2 and Fig. 1.
When ${K_c} > 1$, i.e., the system is in the attractive regime, the RG flow
diagram is the same as Fig. 1 and Fig. 2. Therefor 
we conclude that in this regime
the spin-flip tunneling process is healing the quantum wire.
When $K_c < 1/2$, $J_{2 \perp}$ term is irrelevant and the system is in 
the two-channel Kondo behavior. It reveals from our study that for a certain
region of the initial values of the coupling constants the 
system shows the single
channel Kondo behavior. As we understand from Fig. 3 and Fig.4 that the 
region of two channel Kondo behavior increases as one reduce the value of $K_c$
, i.e., for the higher values of repulsive local potential.
In Fig. 3 and Fig. 4, the region A shows the single channel Kondo behavior and 
the rest of the region of the flow diagram shows the two channel Kondo behavior.
In the two channel Kondo effect there is no healing of quantum wire.
This explicit study of the effect of strong electronic correlations was absent in
the previous study [15]. \\
\section{Conclusions}
We have revisited the problem of magnetic and nonmagnetic impurity in a quantum wire.
We have emphasized mainly the RG study of this problem. The RG flow diagram of
the previous studies is the schematic one [7,12,15]. The Abelian bosonization study only 
reveals that wheather a coupling term is relevant or not. The RG flow diagram
shows us explicitly for which initial values of coupling constants 
the systems flows
to the single channel or two-channel Kondo problem.

\centerline{\bf Acknowledgments}
\vskip .2 true cm
The author would like to acknowledge, Prof. H. R. Krishnamurthy, Prof. Diptiman Sen,
, Prof. P. Durganandini, and Dr. Ravi Chandra for useful discussions
during the progress of the work and also Center for Condensed Matter 
Theory of the 
Physics Department
of IISc for providing working space. 
The author finally acknowledge Dr. B. Mukhopadhay for reading 
the manuscript very critically.

\end{document}